\documentclass[preprint,aps]{revtex4}
\usepackage{comment}
\usepackage{bm} \usepackage{amssymb,amsmath,
amsthm,epsfig,subfigure}  
 \newcommand{\p}{\partial} 
 \makeatletter  \newcommand{\Rmnum}[1]{\expandafter\@slowromancap\romannumeral #1@}

\newcommand{\mx}{\langle x \rangle}
\newcommand{\vx}{{{\rm Var}(x)}}

\usepackage[belowskip=-30pt]{caption}
\makeatother

\begin{document}

\title{Far-from-Equilibrium Time Evolution between two Gamma Distributions}
\author{Eun-jin Kim$^1$, Lucille-Marie Tenk\`es$^{1,2}$, Rainer Hollerbach$^3$,
Ovidiu Radulescu$^4$}
\affiliation{
$^1$School of Mathematics and Statistics,
University of Sheffield, Sheffield, S3 7RH, UK\\
$^2$ENSTA ParisTech Universit\'e Paris-Saclay,
828, boulevard des Mar\'echaux 91120 Palaiseau, France\\
$^3$Department of Applied Mathematics,
University of Leeds, Leeds LS2 9JT, UK\\
$^4$DIMNP-UMR 5235 CNRS, Universit\'e de Montpellier,
Place Eug\`ene Bataillon 34095 Montpellier, France
}
\vspace{1cm}
\begin{abstract}
Many systems in nature and laboratories are far from equilibrium and exhibit
significant fluctuations, invalidating the key assumptions of small fluctuations
and short memory time in or near equilibrium. A full knowledge of Probability
Distribution Functions (PDFs), especially time-dependent PDFs, becomes essential
in understanding far-from-equilibrium processes. We consider a stochastic
logistic model with multiplicative noise, which has gamma distributions as
stationary PDFs. We numerically solve the transient relaxation problem, and show
that as the strength of the stochastic noise increases the time-dependent PDFs
increasingly deviate from gamma distributions. For sufficiently strong noise a
transition occurs whereby the PDF never reaches a stationary state, but instead
forms a peak that becomes ever more narrowly concentrated at the origin. The
addition of an arbitrarily small amount of additive noise regularizes these
solutions, and re-establishes the existence of stationary solutions. In addition
to diagnostic quantities such as mean value, standard deviation, skewness and
kurtosis, the transitions between different solutions are analyzed in terms of
entropy and information length, the total number of statistically distinguishable
states that a system passes through in time.
\end{abstract}

\maketitle

\section{Introduction}
In classical statistical mechanics, the Gaussian (or normal) distribution and
mean-field type theories based on such distributions have been widely used to
describe equilibrium or near equilibrium phenomena. The ubiquity of the Gaussian
distribution stems from the central limit theorem that random variables governed
by different distributions tend to follow the Gaussian distribution in the limit
of large sample size \cite{Fokker,Klebaner,Gardiner}. In such a limit,
fluctuations are small and have a short correlation time, and mean values and
variance completely describe all different moments, greatly facilitating analysis.\\

Many systems in nature and laboratories are however far from equilibrium,
exhibiting significant fluctuations. Examples are found not only in turbulence
in astrophysical and laboratory plasmas, but also in forest fires, the stock
market, and biological ecosystems \cite{nature,KIM02,KIM03,KIM06,KIM08,KIM13,
SrinYoun2011,SayaShowDowl2008,tsuchiya15,tang88,Jensen98,Pruesser12,Longo11,
FLYNN2014,FLYNN2015,Ovidiu,Shahrezaei,Thomas,Biswas,Elgart}. Specifically,
anomalous (much larger than average values) transport associated with large
fluctuations in fusion plasmas can degrade the confinement, potentially even
terminating fusion operation \cite{KIM03}. Tornadoes are rare, large amplitude
events, but can cause very substantial damage when they do occur. 
{\bf
In biology, the pioneering work of Delbr\"uck on bacteriophages showed
that viruses replicate in strongly fluctuating bursts 
\cite{delbruck1945burst}. The fluctuations of the burst 
amplitudes were explained mathematically by stochastic 
autocatalytic reaction models first introduced in \cite{delbruck1940statistical}.
Delbr\"uck's autocatalytic models predict discrete negative-binomial 
distributions, that can be well approximated by gamma distributions when
the average number of particles is large.
}
Furthermore,
gene expression and protein productions, which used to be thought of as smooth
processes, have also been observed to occur in bursts
{\bf leading to
negative binomial and gamma distributed protein copy numbers}
(e.g.\ \cite{Ovidiu,Shahrezaei,Thomas,Biswas,Elgart}). Such rare events of large
amplitude (called intermittency) can dominate the entire transport even if they
occur infrequently \cite{KIM08,KIM09}. They thus invalidate the assumption of
small fluctuations with short correlation time, making mean value and variances
meaningless. Therefore, to understand the dynamics of a system far from
equilibrium, it is crucial to have a full knowledge of Probability Distribution
Functions (PDFs), including time-dependent PDFs \cite{KH16}.\\

{\bf 
The consequences of strong fluctuations in far-from-equilibrium systems are  multiple. In physics, far-from-equilibrium fluctuations produce
dissipative patterns, shift or wipe out phase transitions, etc.  
In economics, finance and actuarial science, strong fluctuations are important issues of risk evaluation. In biology, strong fluctuations generate phenotypic heterogeneity that helps multicellular organisms or microbial populations to adapt to changes of the environment by so-called ``bet-hedging'' strategies. In such a strategy, only a part of the cell population adapts upon emergence of new environmental conditions. The remaining part retains the memory of the old conditions and is thus already
adapted if environmental conditions revert to previous ones
\cite{de2011bet}. Exceptional behavior can also rescue cell subpopulations from drug-induced lethal conditions, thus generating drug resistance \cite{balaban2004bacterial}. In particular, because of the skewness and
exponential tail of the gamma distribution, gamma distributed 
populations contain a significant proportion of individuals with 
exceptionally high phenotype. Bet-hedging being a dynamic phenomenon,
it is important, for biological studies, to be able to predict not
only steady-state but also time-dependent distributions.\\
}

Obtaining a good quality of PDFs is often very challenging, as it requires a
sufficiently large number of simulations or observations. Therefore, a PDF is
usually constructed by averaging data from a long time series, and is thus
stationary (independent of time). Unfortunately, such stationary PDFs miss
crucial information about the dynamics/evolution of non-equilibrium processes
(e.g.\ tumour evolution). Theoretical prediction of time-dependent PDFs has
proven to be no less challenging due to the limitation in our understanding of
nonlinear stochastic dynamical systems as well as the complexity in the required
analysis.\\

Spectral analysis, for example, using theoretical tools similar to those used
in quantum mechanics (e.g.\ raising and lower operators) is useful
(e.g.\ \cite{Fokker}), but the summation of all eigenfunctions is necessary for
time-dependent PDFs far from equilibrium. Various different methodologies have
also been developed to obtain approximate PDFs, such as the variational
principle, the rate equation method, or moment method \cite{prigogine,suzuki80,
suzuki81,langer75,saito76,hasegawa77}. In particular, the rate equation method
\cite{suzuki80,suzuki81} assumes that the form of a time-dependent PDF during
the relaxation is similar to that of the stationary PDF, and thus approximates
a time-dependent PDF during transient relaxation by a PDF having the same
functional form as a stationary PDF, but with time-varying parameters.\\

In this work we show that this assumption is not always appropriate. We consider
a stochastic logistic model with multiplicative noise. We show that for fixed
parameter values the stationary PDFs are always gamma distributions
(e.g.\ \cite{dennis,liao}), one of the most popular distributions used in
fitting experimental data. However, we find numerically that the time-dependent
PDFs in transitioning from one set of parameter values to another are
significantly different from gamma distributions, especially for strong
stochastic noise. For sufficiently strong multiplicative noise it is necessary
to introduce additive noise as well to obtain stationary distributions at all.
{\bf 
We note that in inferential statistics, gamma distributions facilitate Bayesian model learning from data, 
as a gamma distribution is a conjugate prior to many likelihood 
functions. It is therefore interesting to test whether models with
stationary gamma distributions also have time-dependent gamma 
distributions.}

\section{Stochastic Logistic Model}
We consider the logistic growth with a multiplicative noise given by the
following Langevin equation:
\begin{equation}
\frac{dx}{dt} =   (\gamma + \xi) x - \epsilon x^2,\label{eq2}
\end{equation}
where $x$ is a random variable, and $\xi$ is a stochastic forcing, which for
simplicity can be taken as a short-correlated random forcing as follows:
\begin{equation}
\langle \xi(t) \xi(t') \rangle={ 2D} \delta (t - t').\label{eq3}
\end{equation}
In Eq.\ (\ref{eq3}), the angular brackets represent the average over $\xi$,
$\langle{\xi} \rangle=0$, and $D$ is the strength of the forcing. $\gamma$ is
the control parameter in the positive feedback, representing the growth rate
of $x$, while $\epsilon$ represents the efficiency in self-regulation by a
negative feedback. 
{\bf
$\gamma x - \epsilon x^{2}$ can be 
considered as the gradient of the potential $V$ as
$ \gamma x - \epsilon x^{2} = - \frac{\partial V}{\partial x},$
where $V = -\frac{\gamma}{2} x^{2} + \frac{\epsilon}{3} x^{3 }$. 
Thus, $V$ has its minimum value at $x=\frac{ \gamma}{ \epsilon}$.
When $\xi=0$,
$x=\frac{\gamma}{\epsilon}$ (the carrying capacity) is a stable equilibrium point since 
$\partial_{{xx}} V \Bigr|_{x={\gamma }/{\epsilon}} = \gamma >0$; 
$x=0$ is an unstable equilibrium point since $\partial_{xx} V \Bigr| _{{x=0}} = -\gamma < 0$. 
}
The multiplicative noise in Eq.\ (\ref{eq2}) shows that the linear growth rate
contains the stochastic part $\xi$.
{\bf
This model is entirely phenomenological and $x$ can be interpreted as
the size of a critical physical phenomenon (vortex, tornado, etc.),
stock market, number of biological cells, viruses, proteins. 
It is reminiscent of Delbr\"uck's autocatalytic processes \cite{delbruck1940statistical}, but is different
from these in many ways, the most important being the lack of 
discreteness and the possibility of reaching a steady state due to
the finite capacity of logistic growth. We will show in the 
following that in spite of these differences, our model is capable of producing large fluctuations. \\
}

By using the Stratonovich calculus \cite{Klebaner,Gardiner,WongZakai}, we can
obtain the following Fokker-Planck equation for the PDF $p(x,t)$ (see Appendix
A for details):
\begin{equation}
\frac{\p}{\p t}\,p(x,t)
= - \frac{\p}{\p x}\Bigl[(\gamma x -\epsilon x^2)p(x,t)\Bigr]
+D\,\frac{\p}{\p x}\left[x\,\frac{\p}{\p x}\Bigl[x\,p(x,t)\Bigr]\right]
\label{eq4}
\end{equation}
corresponding to the Langevin equation (\ref{eq2}). By setting
$\partial_t p=0$, we can analytically solve for the stationary PDFs as
\begin{equation}
p(x) = \frac{b^a}{\Gamma(a) } x^{{a-1}} e^{-b x},
 \label{eq5}
\end{equation}
which is the well-known gamma distribution. The two parameters $a$ and $b$ are
given by $a={\gamma}/{D}$ and $b={\epsilon}/{D}$. The mean value and variance of
the gamma distribution are found to be:
\begin{equation}
\langle x \rangle=\frac{a}{b}=\frac{\gamma}{\epsilon},\qquad\qquad
{\rm Var}(x)=\sigma^{2}=\langle (x-\langle x \rangle )^2\rangle=\frac{a}{b^2}
=\frac{\gamma D}{\epsilon^2},
\label{eq6}
\end{equation}
where $\sigma =\sqrt{{\rm Var}(x)}$ is the standard deviation. We recognise
$\langle x \rangle $ as the carrying capacity for a deterministic system with
$\xi=0$. It is useful to note that $\mx$ is given by the linear growth rate
scaled by $\epsilon$, while $\vx$ is given by the product of the linear growth
rate and the diffusion coefficient, each scaled by $\epsilon$. That is, the
effect of stochasticity should be measured relative to the linear growth rate.\\

Therefore, the case of small fluctuations is modelled by values of $D$ small
compared with $\gamma$ and $\epsilon$. In such a limit, $a$ and $b$ are large,
making $\sqrt{{\rm Var}(x)} \ll \langle x \rangle$ in Eq.\ (\ref{eq6}). That is,
the width of the PDF is much smaller than its mean value. In this limit,
Eq.\ (\ref{eq5}) reduces to a Gaussian distribution. To show this, we express
Eq.\ (\ref{eq5}) in the following form:
\begin{equation}
p \equiv \frac{b^a}{\Gamma(a) } e^{-f(x)},
\label{eq7}
\end{equation}
where $f(x) = bx - (a-1)\ln{x}$. For large $b$, we expand $f(x)$ around the
stationary point $x=x_p$ where $\partial_x f(x)=0= b-(a-1)/x$ up to the second
order in $x-x_p$ to find: 
\begin{eqnarray}
&&x_p = \frac{a-1}{b} \sim \frac{a}{b},\qquad\qquad f(x=x_p)
 \sim a \left( 1 - \ln{\frac{a}{b}} \right),
\label{eq8}\\
&& f(x) \sim f(x_p) + \frac{1}{2}  (x-x_p)^2 \partial_{xx} f(x)\Big|_{x=x_p}
=a\left(1-\ln{\frac{a}{b}}\right)+\frac{b^2}{2 a}\left(x-\frac{a}{b}\right)^2.
\label{eq9}
\end{eqnarray} 
Here $a \gg 1$ was used. Using Eq.\ (\ref{eq9}) in Eq.\ (\ref{eq7}) then gives
us
\begin{equation}
p\propto \exp{\left[-\frac{b^2}{2 a} \left ( x - \frac{a}{b} \right)^2\right]}
\propto \exp{\Bigl[-\beta (x-\langle x \rangle)^2 \Bigr]},
\label{eq10}
\end{equation}
which is a Gaussian PDF with mean value $\langle x \rangle$. Here $\beta=
{1}/{{\rm Var}(x)}$ is the inverse temperature and the variance ${\rm Var}(x)$
is given by Eq.\ (\ref{eq6}). Therefore, for a sufficiently small $D$, the gamma
distribution is approximated as a Gaussian PDF, which is consistent with the
central limit theorem as small $D$ corresponds to small fluctuations and large
system size. See also \cite{bagui} for a different derivation.\\

As $D$ increases, the Gaussian approximation becomes increasingly less valid.
Indeed, even the gamma distribution becomes invalid asymptotically, when $t \to \infty$, if $D > \gamma$;
according to Eq.\ (\ref{eq5}) having $a<1$ yields $\displaystyle\lim_{x\to0}p
=\lim_{x\to0}\frac{\partial p}{\partial x}=\infty$.
However, from the full time-dependent Fokker-Planck equation
(\ref{eq4}) one finds that if the initial condition satisfies $p=0$ at $x=0$,
then $p(x=0)$ will remain 0 for all later times. As we will see, the resolution
to this seeming paradox is that no stationary distribution is ever reached for
$D>\gamma$, but instead the peak approaches ever closer to $x=0$, without ever
reaching it.\\

If we are interested in obtaining stationary solutions even when $D>\gamma$, one
way to achieve that is to return to the original Langevin equation (\ref{eq2}),
but now include a further additive stochastic noise $\eta$:
\begin{equation}
\frac{dx}{dt} =   (\gamma + \xi) x - \epsilon x^2 + \eta,\label{L2}
\end{equation}
where $\xi$ and $\eta$ are uncorrelated, and $\eta$ satisfies $\langle\eta(t)
\eta(t')\rangle=2Q\delta(t-t')$. The new version of the Fokker-Planck
equation (\ref{eq4}) then becomes:
\begin{equation}
\frac{\p}{\p t}\,p
= - \frac{\p}{\p x}\Bigl[(\gamma x -\epsilon x^2)p\Bigr]
+D\,\frac{\p}{\p x}\left[x\,\frac{\p}{\p x}\Bigl[x\,p\Bigr]\right]
+Q\frac{\p^2}{\p x^2}\,p,
\label{FP2}
\end{equation}
which has stationary solutions given by
\begin{equation}
\ln p(x)=\int\frac{(\gamma-D)x-\epsilon x^2}{Dx^2+Q}\,dx.
\label{fp2stationary}
\end{equation}
This integral can be evaluated analytically, but the final form is not
particularly illuminating. The only point to note is that for non-zero $Q$
the denominator is never 0 even for $x\to0$, which avoids any possible
singularities at the origin. For $\gamma>D$ and $Q\ll D$ the solutions are
also essentially indistinguishable from the previous gamma distribution
(\ref{eq5}). The only significant effect of including $\eta$ therefore is to
avoid the previous difficulties at the origin when $D>\gamma$.\\

As we have seen, both Fokker-Planck equations (\ref{eq4}) and (\ref{FP2}) can
be solved exactly for their stationary solutions. This is unfortunately not the
case regarding time-dependent solutions, where no closed-form analytic solutions
exist. (See Appendix B for the extent to which analytic progress can be made.)
We therefore developed finite-difference codes, second-order accurate in both
space and time. Most aspects of the numerics are standard, and similar to
previous work \cite{KH17,Entropy,paper6}. The only point that requires
discussion are the boundary conditions. As noted above, for (\ref{eq4}) the
equation itself states that $p=0$ at $x=0$ is the appropriate boundary
condition, provided only that the initial condition also satisfies this. In
contrast, for (\ref{FP2}) the appropriate boundary condition is $\frac{\p}{\p x}
p=0$ at $x=0$. To derive this boundary condition for (\ref{FP2}),
 we simply integrate (\ref{FP2}) over the range
$x=[0,\infty]$ and require that the total probability should always remain 1,
so that $\frac{d}{dt}\int p\,dx=0$. Regarding the outer boundary, choosing some
moderately large outer value for $x$, and then imposing $p=0$ there was
sufficient. Resolutions up to $10^6$ grid points were used, and results were
carefully checked to ensure they were independent of the grid size, time step,
and precise choice of outer boundary.\\

\section{Diagnostics}
Once the time-dependent solutions are computed, we can analyze them using a
number of diagnostics. First, we can evaluate the mean value $\langle x\rangle$
and standard deviation $\sigma$ from (\ref{eq6}). Next, to explore the extent to
which the time-dependent PDFs differ from gamma distributions, we can simply
compare them with `equivalent' gamma distributions and compute the difference.
That is, given $\langle x\rangle$ and $\sigma$, the gamma distribution
$p_{\rm equiv}$ having the same mean and variance would have as its two
parameters $a=\langle x\rangle^2/\sigma^2$ and $b=\langle x\rangle/\sigma^2$.
With these values, we define
\begin{equation}
{\rm Difference}=\int|p - p_{\rm equiv}|\,dx
\label{difference}
\end{equation}
to measure how different the actual time-dependent PDF is from its equivalent
gamma distribution.\\

Two other familiar quantities often useful in analyzing PDFs are the skewness
and kurtosis, defined by
\begin{equation}
{\rm Skewness}=\frac{\langle(x-\langle x\rangle)^3\rangle}{\sigma^3},\qquad
{\rm Kurtosis}=\frac{\langle(x-\langle x\rangle)^4\rangle}{\sigma^4}-3.
\label{skewkurt}
\end{equation}
Skewness measures the extent to which a PDF is asymmetric about its peak,
whereas kurtosis measures how concentrated a PDF is in the peak versus the
tails, relative to a Gaussian having the same variance. (The $-3$ is included
in the definition of the kurtosis to ensure that a Gaussian would yield 0.)
For gamma distributions one finds analytically that the skewness is
$2\sqrt{D/\gamma}$, and the kurtosis is $6D/\gamma$. Comparing the skewness
and kurtosis of the time-dependent PDFs with these formulas is therefore
another useful way of quantifying how similar or different they are from
gamma distributions.\\

Another quantity that can be useful is the so-called differential entropy
as a measure of order versus disorder (as entropy always is):
\begin{equation}
S=-\int p\ln p\,dx.
\label{entropy}
\end{equation}
In particular, we expect $S$ to be small for localised PDFs, and large for spread
out ones (e.g.\ \cite{KH17,Entropy,paper6,fisher}). For unimodal PDFs as the
ones studied here, entropy and standard deviation are typically comparably good
measures of localization, but for bimodal peaks entropy can be significantly
better \cite{paper6}.
{\bf
For the gamma distribution in Eq.\ (\ref{eq5}),
 the differential entropy can be shown to be given by
\begin{equation}
S= a - \ln{b} + \ln{(\Gamma(a))} + (1-a) \psi(a),
\label{entropy}
\end{equation}
where $\psi(a)=\frac{d \ln{(\Gamma(x))}}{dx}  \bigr|_{x=a}$ is the double gamma function. \\
}

Our final diagnostic quantity is what is known as {\it information length}.
Unlike all the previous diagnostics, which are simply evaluated at any instant
in time but otherwise do not involve $t$, information length is the Lagrangian
quantity, explicitly concerned with the full time-history of the evolution of a given PDF. 
It is thus ideally suited to understanding time-dependent PDFs. Very briefly, we begin by
defining
\begin{equation}
{\cal{E}} \equiv \frac{1}{[\tau (t)]^2} =
 \int \frac {1} {p(x,t)}\left[\frac{\partial p(x,t)}{\partial t}\right]^2\,dx.
\label{curlyE}
\end{equation}
Note how $\tau$ has units of time, and quantifies the correlation time over
which the PDF changes, thereby serving as a time unit in statistical space.
Alternatively, $1/\tau$ quantifies the (average) rate of change of information
in time. 
{\bf 
${\cal E}$ is due to the change in either width (variance) of the PDF 
or the mean value, which are determined by $\gamma$, $D$ and $\epsilon$ 
for the gamma distribution (e.g.\ see Eq.\ (\ref{eq5})). In the standard Brownian motion, 
the mean value is zero so that ${\cal E}$ is due to the change in the variance of a PDF.}\\

The total change in information between initial and final times, $0$ and $t$
respectively, is then defined by measuring the total elapsed time in units of
$\tau$ as:
\begin{equation}
{\cal{L}} (t) = \int_0^{t} \frac{dt_1}{\tau(t_1)}
 = \int_0^{t} \sqrt{\int dx \frac {1} {p(x, t_1)}
 \left [\frac {\partial p(x,t_1)} {\partial t_1} \right]^2}\,dt_1.
\label{curlyL}
\end{equation}
This information length ${\cal L}$ measures the total number of statistically
distinguishable states that a system evolves through, thereby establishing a
distance between the initial and final PDFs in the statistical space. 
{\bf
Note that ${\cal L}$ by construction is a continuous variable, and thus measures 
the total `number' of statistically different states as a continuous number.}
See also \cite{fisher,WOOTTERS81,NK14,NK15,HK16,KIM16,KH17,Entropy,paper6} for
further applications and theoretical background of ${\cal E}$ and ${\cal L}$.\\

\section{Results}

\subsection{$\gamma>D$}
We start with the case $\gamma>D$, where Eq.\ (\ref{eq4}) has stationary
solutions, given by (\ref{eq5}). Keeping $\epsilon$ and $D$ fixed, we then
switch $\gamma$ back and forth between two values, in the following sense:
Take the gamma distribution (\ref{eq5}) corresponding to one value, call it
$\gamma_1$, and use that as the initial condition to solve (\ref{eq4}) with
the other value, call it $\gamma_2$. We then interchange $\gamma_1$ and
$\gamma_2$ to complete the pair of `inward' and `outward' processes. 
Such a
pair can be thought of as an order/disorder phase transition
\cite{KH17,Entropy}, caused for example by cyclically adjusting temperature
in an experiment.
{\bf 
This protocol is also inspired from adaptation
of a biological system. During adaptation a model parameter can be abruptly
changed in response to the change of environmental conditions, for 
instance a particle replication parameter $\gamma$, but the resulting
changes can be extremely heterogeneous in the population.
}

\begin{figure}
\centering
\includegraphics[scale=0.9]{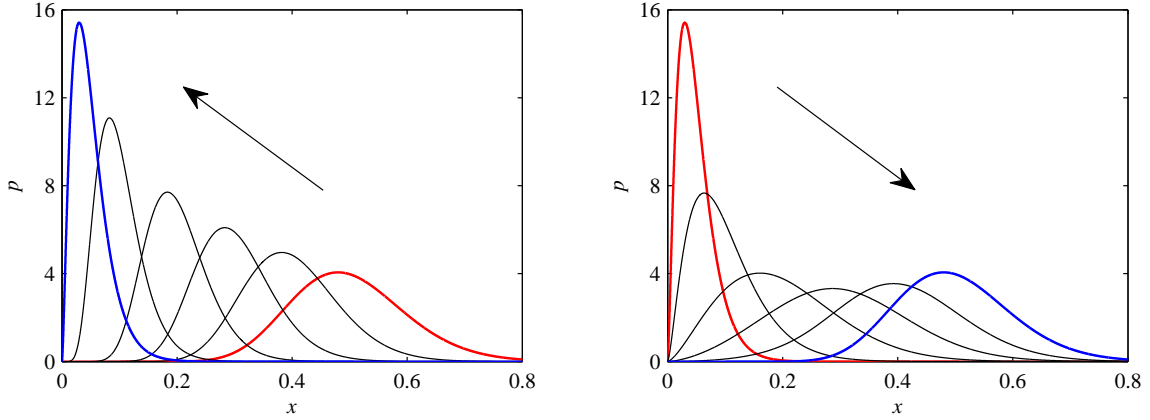}
\caption{The left panel shows the result of switching $\gamma=0.5\to0.05$,
the right panel $\gamma=0.05\to0.5$, both at fixed $\epsilon=1$ and $D=0.02$.
The initial (red) and final (blue) gamma distributions are shown as heavy lines.
The four intermediate lines are when the time-dependent solutions have
$\langle x\rangle=0.1,\ 0.2,\ 0.3,\ 0.4$. The arrows are a reminder of the
direction of motion, inward on the left and outward on the right.}
\vskip 1cm
\label{Fig1}
\end{figure}

Figure \ref{Fig1} shows the result of switching $\gamma$ between $\gamma_1=0.5$
and $\gamma_2=0.05$, for fixed $\epsilon=1$ and $D=0.02$. (One of the three
parameters $\epsilon$, $\gamma$ and $D$ can of course always be kept fixed by
rescaling the entire equation, so throughout this entire section we keep
$\epsilon=1$ fixed, and focus on how the various quantities depend on $\gamma$
and $D$.) We immediately see that the inward and outward processes
behave differently. When $\gamma$ is decreased, and the peak therefore moves
inward, the PDF is relatively narrow, and the peak amplitude is monotonically
increasing. When $\gamma$ is switched from 0.05 back to 0.5, the PDF is much
broader, and the peak amplitude in the intermediate stages is less than either
the initial or final gamma distributions.\\

\begin{figure}
\centering
\includegraphics[scale=0.9]{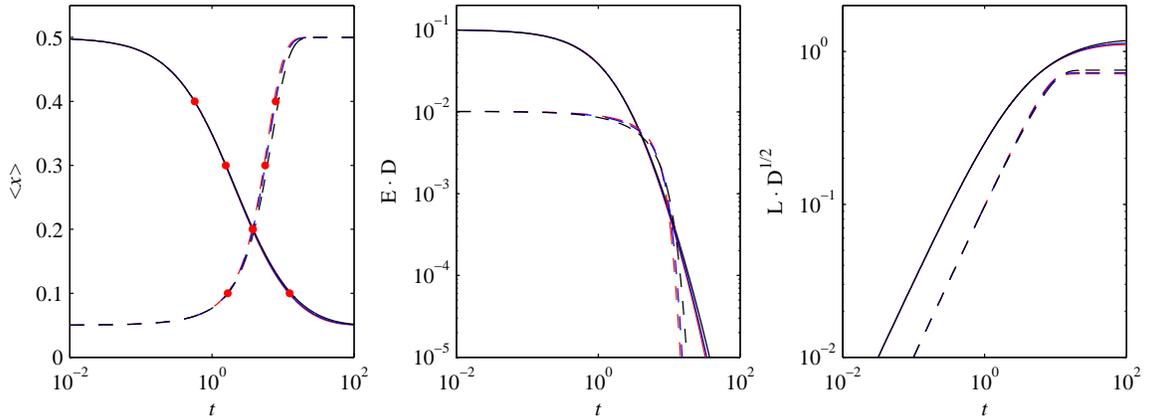}
\caption{The first panel shows $\langle x\rangle$ as a function of time, the
second panel shows ${\cal E}\cdot D$ (to indicate the ${\cal E}\sim D^{-1}$
scaling), and the third panel shows ${\cal L}\cdot D^{1/2}$ (to indicate the
${\cal L}\sim D^{-1/2}$ scaling). Solid lines denote $\gamma=0.5\to0.05$,
dashed lines the reverse. Each solid or dashed `line' is in fact three --
occasionally barely distinguishable -- lines with $D=0.01,\ 0.02,\ 0.04$.
The dots on the $\langle x\rangle$ curves correspond to the PDFs shown in
figure \ref{Fig1}.}
\vskip 1cm
\label{Fig2}
\end{figure}

Figure \ref{Fig2} shows how $\langle x\rangle$, ${\cal E}$ and ${\cal L}$ vary
as functions of time, for the three values $D=0.01,\ 0.02,\ 0.04$. For
$\langle x\rangle$ the movement from 0.5 to 0.05 is somewhat slower than the
reverse process, but both processes occur on a similar timescale, and both are
essentially independent of $D$. This is in contrast with other Fokker-Planck
systems where the magnitude of the diffusion coefficient can have a very strong
influence on the equilibration timescales \cite{KH17,Entropy}.\\

For $\cal E$ and $\cal L$ the equilibration is again somewhat slower for
$\gamma=0.5\to0.05$ than the reverse. We can further identify clear scalings
${\cal E}\sim D^{-1}$ and ${\cal L}\sim D^{-1/2}$. Finally, $\cal L$ is greater
for $\gamma=0.5\to0.05$ than the reverse. These results are all understandable
in terms of the interpretation of $\cal L$ as the number of statistically
distinguishable states that the PDF evolves through: First, we recall from
figure \ref{Fig1} that $\gamma=0.5\to0.05$ had consistently narrower PDFs than
the reverse. Narrower PDFs means more distinguishable states, hence larger
$\cal L$ for $\gamma=0.5\to0.05$ than the reverse. The ${\cal L}\sim D^{-1/2}$
scaling has the same explanation; smaller $D$ yields narrower PDFs, hence
larger $\cal L$.\\

\begin{figure}
\centering
\includegraphics[scale=0.9]{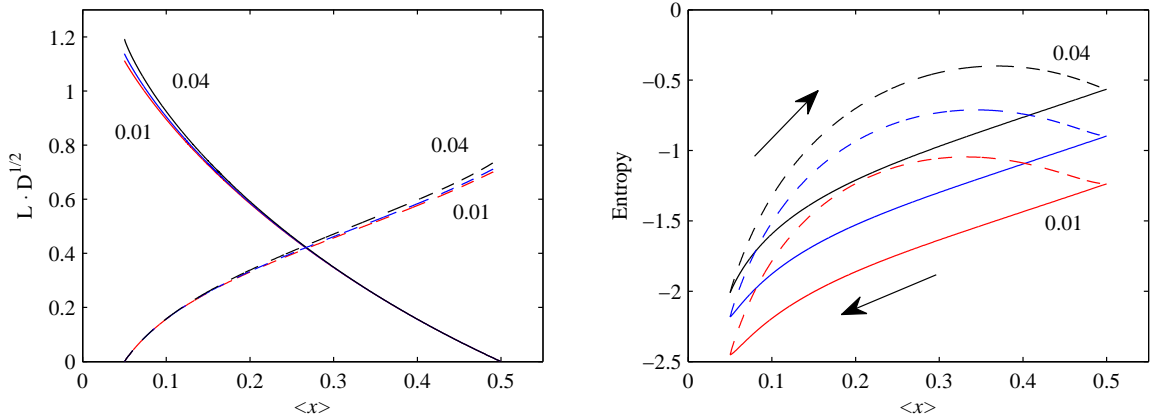}
\caption{The left panel shows ${\cal L}\cdot D^{1/2}$, the right panel entropy,
both as functions of $\langle x\rangle$. Solid lines denote $\gamma=0.5\to0.05$,
dashed lines the reverse. Numbers besides curves indicate $D=0.01,\ 0.02,\ 0.04$.
The arrows on the entropy plot are a reminder of the direction of inward/outward
motion.}
\vskip 1cm
\label{Fig3}
\end{figure}

The first panel in figure \ref{Fig3} shows the previous quantities $\langle x
\rangle$ and ${\cal L}\cdot D^{1/2}$, but now plotted against each other rather
than separately against time. The behaviour is exactly as one might expect, with
$\cal L$ growing more or less linearly with distance from the initial position.
The right panel in figure \ref{Fig3} shows the entropy (\ref{entropy}), again as
a function of $\langle x\rangle$ rather than time, to emphasize the cyclic
nature of the two processes. The significance is indeed as claimed above, with
more localized PDFs having smaller entropy values. Note how $\gamma=0.5\to0.05$,
which had the narrower PDFs, has lower entropy values than the reverse process.
Note also how reducing $D$ by a factor of two, thereby making the PDFs narrower,
causes the entire cyclic pattern to shift downward by an essentially constant
amount.\\

\begin{figure}
\centering
\includegraphics[scale=0.9]{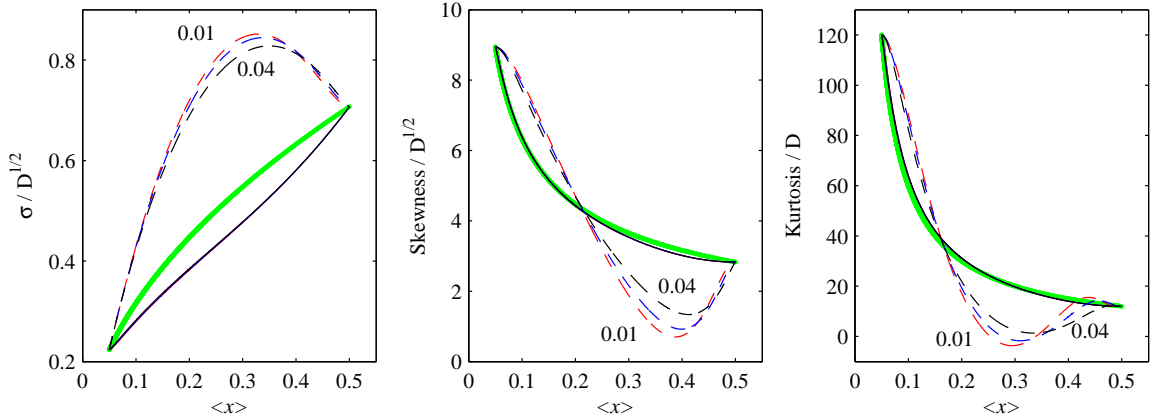}
\caption{$\sigma/D^{1/2}$, (skewness$\,/D^{1/2}$) and (kurtosis$\,/D$), as
functions of $\langle x\rangle$. Solid lines denote $\gamma=0.5\to0.05$,
dashed lines the reverse. Numbers besides curves indicate $D=0.01,\ 
0.02,\ 0.04$. The heavy green curves are $\sqrt{\langle x\rangle}$,
$2/\sqrt{\langle x\rangle}$ and $6/{\langle x\rangle}$, respectively,
and indicate the behaviour expected for exact gamma distributions.}
\vskip 1cm
\label{Fig4}
\end{figure}

Figure \ref{Fig4} shows how the standard deviation, skewness and kurtosis
behave, again as functions of $\langle x\rangle$ throughout the two processes.
The heavy green lines also show the behaviour that would be expected if the
time-dependent PDFs were always gamma distributions throughout their evolution.
That is, if gamma distributions have $\langle x\rangle=\gamma$, $\sigma=
\sqrt{\gamma D}$, skewness$\;=2\sqrt{D/\gamma}$ and kurtosis$\;=6D/\gamma$
(setting $\epsilon=1$), then expressed as function of $\langle x\rangle$ we
would have $\sigma/D^{1/2}=\sqrt{\langle x\rangle}$, (skewness$\,/\sqrt{D})=2/
\sqrt{\langle x\rangle}$ and (kurtosis$\,/D)=6/{\langle x\rangle}$. As we can
see, the $\gamma=0.5\to0.05$ process follows these functional relationships
reasonably well (especially for skewness and kurtosis), but for
$\gamma=0.05\to0.5$ all three quantities deviate substantially.\\

\begin{figure}
\centering
\includegraphics[scale=0.9]{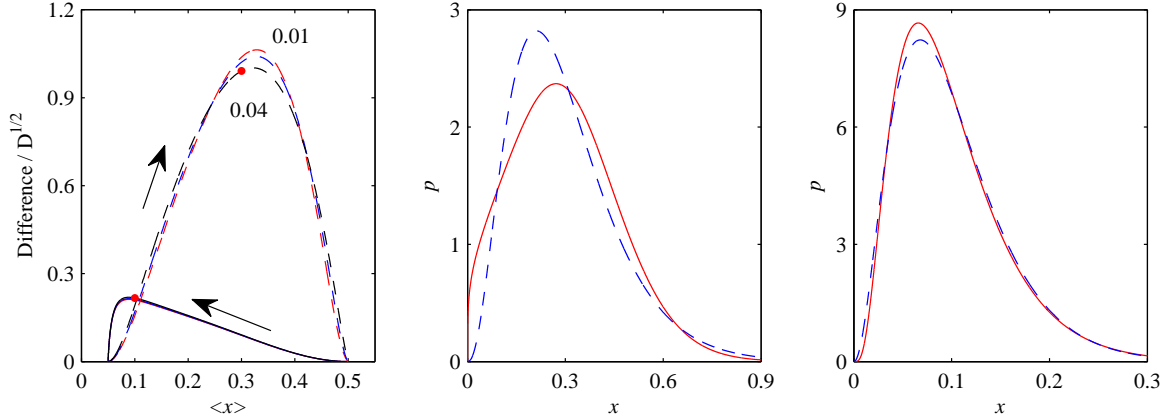}
\caption{The first panel shows the difference (\ref{difference}) between
the actual PDF and the equivalent gamma distribution, as functions of
$\langle x\rangle$. Solid lines denote $\gamma=0.5\to0.05$, dashed lines
the reverse, with arrows also indicating the direction of motion. The dots at
$\langle x\rangle=0.3$ for $\gamma=0.05\to0.5$, and $\langle x\rangle=0.1$ for
$\gamma=0.5\to0.05$, correspond to the other two panels: Panel 2 compares
the $\gamma=0.05\to0.5$ PDF with its equivalent gamma distribution; Panel 3
compares the $\gamma=0.5\to0.05$ PDF with its equivalent gamma distribution.
The actual PDFs in each case are solid (red), and the equivalent gamma
distributions are dashed (blue). $D=0.04$ for both sets.}
\vskip 1cm
\label{Fig5}
\end{figure}

Further evidence of significant deviations from gamma distribution behaviour
is seen in figure \ref{Fig5}, showing the difference (\ref{difference})
directly. As expected from figure \ref{Fig4}, $\gamma=0.05\to0.5$ has a much
greater difference than $\gamma=0.5\to0.05$. The second and third panels show
how the PDFs compare with the equivalent gamma distributions having the same
$\langle x\rangle$ and $\sigma$ values as the actual PDFs at that instant. The
differences are clearly visible, especially for $\gamma=0.05\to0.5$, but also
for $\gamma=0.5\to0.05$.\\

\subsection{$D>\gamma$}
We next consider the case $D>\gamma$, where we demonstrated above that
stationary solutions cannot exist at all, because the time-dependent PDF can
only ever get closer and closer to the gamma distribution singularity at the
origin, but can never actually achieve it. To explore what does happen in this
case then, we simply repeat the above procedure, except that there is now only
an `inward' process, and no reverse. That is, instead of $\gamma=0.5\to0.05$,
let us consider $\gamma=0.5\to0$. (Throughout this section we will also take
$D=10^{-3}$, to facilitate comparison with results in the next section. For
$\gamma=0$ of course any $D$ is greater than $\gamma$.)\\

\begin{figure}
\centering
\includegraphics[scale=0.9]{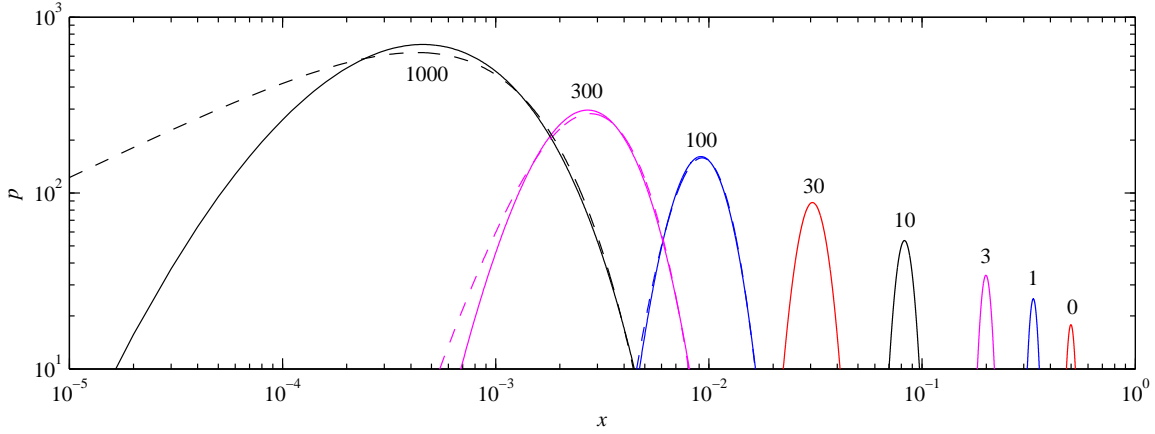}
\caption{The initial condition is a gamma distribution with $\gamma=0.5$,
$\epsilon=1$ and $D=10^{-3}$; $\gamma$ is then switched to 0, and the solution
is evolved according to Eq.\ (\ref{eq4}). Numbers besides curves indicate time,
from the initial condition at $t=0$ to the final time 1000. The dashed curves
indicate the equivalent gamma distributions having the same $\langle x\rangle$
and $\sigma$.}
\vskip 1cm
\label{Fig6}
\end{figure}

Figure \ref{Fig6} shows the resulting PDFs, and how they approach ever closer
to the origin, but never actually achieve the $x^{-1}$ blowup that would be
implied by Eq.\ (\ref{eq5}) for $a=\gamma/D=0$. The peak amplitude simply
increases indefinitely, as $t^{1/2}$. The widths correspondingly also decrease;
the apparent increase is an illusion caused by the logarithmic scale for $x$.
The dashed lines also show the equivalent gamma distributions, as before. Note
how the difference becomes increasingly noticeable; in line with the fact that
the equivalent gamma distribution is tending toward its singular behaviour as
$\langle x\rangle$ decreases, but the actual PDFs must always have $p(0)=0$.\\

\begin{figure}
\centering
\includegraphics[scale=0.9]{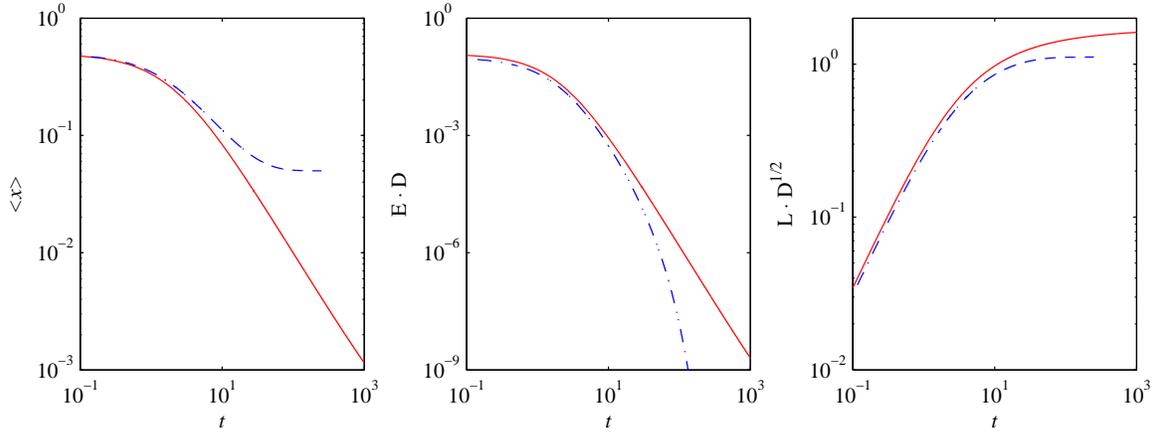}
\caption{As in figure \ref{Fig2}, the first panel shows $\langle x\rangle$, the
second panel shows ${\cal E}\cdot D$, and the third panel ${\cal L}\cdot D^{1/2}$.
Solid lines denote $\gamma=0.5\to0$ for $D=10^{-3}$, dashed lines the previous
$\gamma=0.5\to0.05$ for $D=0.01$. Note how the scalings of $\cal E$ and $\cal L$
with $D$ are still preserved even when $D$ is changed by a factor of 10.}
\vskip 1cm
\label{Fig7}
\end{figure}

Figure \ref{Fig7} is the equivalent of figure \ref{Fig2}, and directly compares
$\gamma=0.5\to0$ here with the previous $\gamma=0.5\to0.05$. We see that
$\langle x\rangle$ starts out very similarly, but instead of equilibrating to
0.05, it now tends to 0 as $t^{-1}$. $\cal E$ again starts out similarly, but
ultimately tends to 0 much slower, as $t^{-3}$ instead of exponentially. This
$t^{-3}$ scaling for $\cal E$ has an interesting consequence for $\cal L$,
namely that $\cal L$ does saturate to a finite value ${\cal L}_\infty$ (since
$\int t^{-3/2}\,dt$ remains bounded for $t\to\infty$) even though the PDF
itself never settles to a stationary state.\\

\begin{figure}
\centering
\includegraphics[scale=0.9]{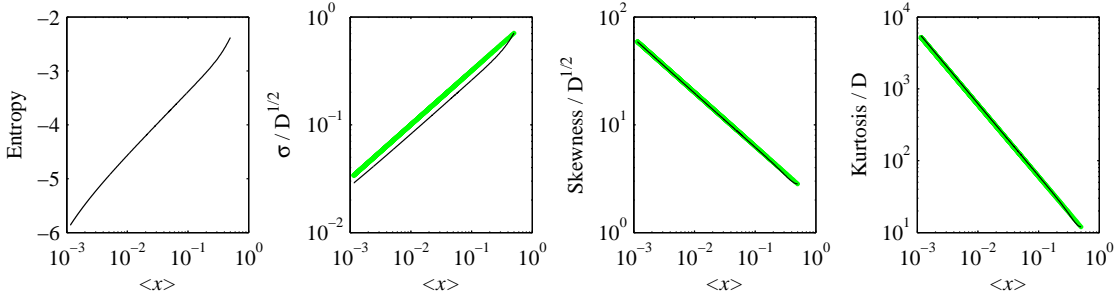}
\caption{Entropy, $\sigma/D^{1/2}$, (skewness$\,/D^{1/2}$) and (kurtosis$\,/D$),
as functions of $\langle x\rangle$, for the $\gamma=0.5\to0$ calculation from
figure \ref{Fig6}. The heavy green curves in the last three panels are
$\sqrt{\langle x\rangle}$, $2/\sqrt{\langle x\rangle}$ and $6/{\langle
x\rangle}$, respectively, and indicate the behaviour expected for exact gamma
distributions.}
\vskip 1cm
\label{Fig8}
\end{figure}

Figure \ref{Fig8} shows entropy, $\sigma$, skewness and kurtosis, so some of
the results as in figures \ref{Fig3} and \ref{Fig4}. Entropy and $\sigma$ are
again both good measures of how narrow the PDF is, becoming ever smaller as the
peak moves toward the origin. Skewness and kurtosis seem to follow the expected
gamma distribution relationship extremely well, even though we saw before in
figure \ref{Fig6} that the PDFs are actually different from gamma distributions.
As $\langle x\rangle\to0$, both skewness and kurtosis thus become indefinitely
large.\\

\subsection{$Q\neq0$}
Finally, we turn to the Fokker-Planck equation (\ref{FP2}) with additive
noise included, and use it to explore the two questions that could not be
addressed otherwise. First, how does a process like $\gamma=0.5\to0$ then
equilibrate to a stationary solution? Second, what does the reverse process
$\gamma=0\to0.5$ look like?\\

We will keep $D=10^{-3}$ and $Q=10^{-5}$ fixed throughout this section.
Since the effective diffusion coefficients in (\ref{FP2}) are $Dx^2$ and $Q$
[recall also the denominator of Eq.\ (\ref{fp2stationary})], this means that
$Q$ is dominant only within $x\le0.1$; any stationary solutions with peaks
much beyond that are effectively pure gamma distributions.\\

\begin{figure}
\centering
\includegraphics[scale=0.9]{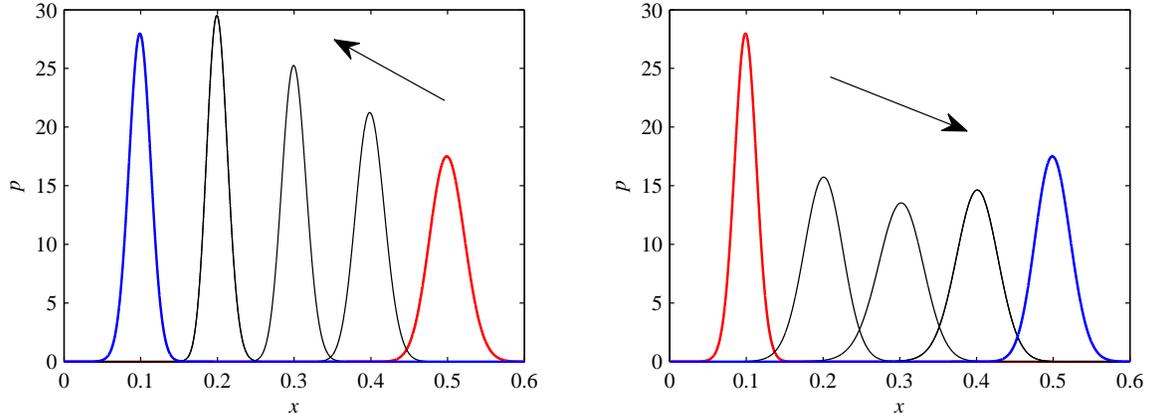}
\caption{The left panel shows the result of switching $\gamma=0.5\to0.1$,
the right panel $\gamma=0.1\to0.5$, both at fixed $\epsilon=1$, $D=10^{-3}$ and
$Q=10^{-5}$. The initial (red) and final (blue) gamma distributions are shown
as heavy lines. The three intermediate lines are when the time-dependent
solutions have $\langle x\rangle=0.2,\ 0.3,\ 0.4$. ${\cal L}_\infty=25$ on the
left and 16 on the right.}
\vskip 1cm
\label{Fig9}
\end{figure}

Figure \ref{Fig9} shows the same type of inward/outward process as before
in figure \ref{Fig1}, only now switching $\gamma$ between 0.5 and 0.1.
Comparing with figure \ref{Fig1}, we see that the dynamics are very similar,
just with all the peaks considerably narrower, which is to be expected if
$D=10^{-3}$ rather than 0.02. The only other point to note is how the final
peak in the left panel is lower than the previous peak at $\langle x\rangle=0.2$,
which is different from figure \ref{Fig1}, where $\gamma=0.5\to0.05$ had peaks
monotonically increasing throughout the entire evolution. The reason the final
peak here decreases slightly is precisely the influence of $Q$ in this region;
if this peak is now seeing just as much diffusion from $Q$ as from $D$, it is
not surprising that it spreads out somewhat more, and is correspondingly
somewhat lower than a pure gamma distribution would be.\\

\begin{figure}
\centering
\includegraphics[scale=0.9]{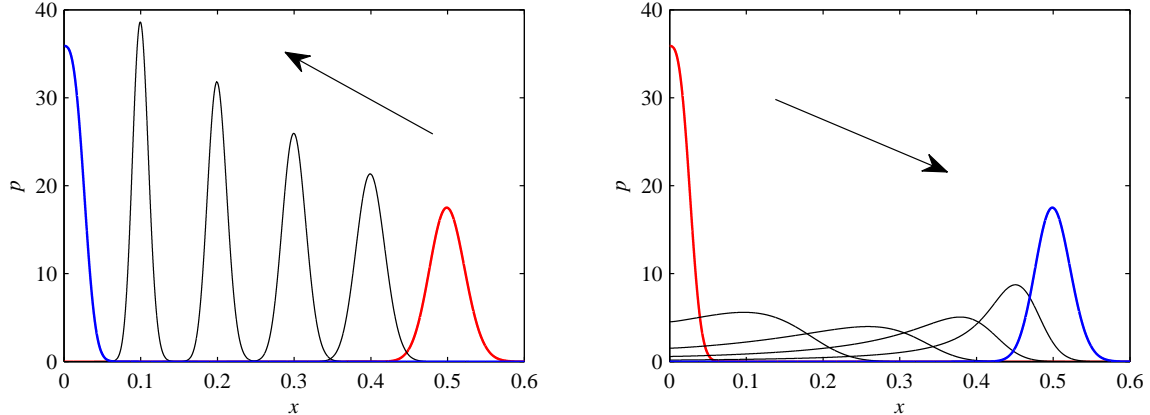}
\caption{The left panel shows the result of switching $\gamma=0.5\to0$,
the right panel $\gamma=0\to0.5$, both at fixed $\epsilon=1$, $D=10^{-3}$ and
$Q=10^{-5}$. The initial (red) and final (blue) gamma distributions are shown
as heavy lines. The four intermediate lines are when the time-dependent
solutions have $\langle x\rangle=0.1,\ 0.2,\ 0.3,\ 0.4$. ${\cal L}_\infty=35$
on the left and 9.5 on the right.}
\vskip 1cm
\label{Fig10}
\end{figure}

Figure \ref{Fig10} shows the fundamentally new case, namely switching
$\gamma$ between 0.5 and 0. The inward process $\gamma=0.5\to0$ is again very
similar to either figure \ref{Fig1} or \ref{Fig9}. The only difference to figure
\ref{Fig6} is that the process does actually equilibrate to a stationary solution
now, as given by Eq.\ (\ref{fp2stationary}). The reverse process $\gamma=0\to0.5$
is rather different though. The initial central peak now broadens far more
than previously seen in figures \ref{Fig1} and \ref{Fig9}.\\

One interesting consequence of this extreme broadening for $\gamma=0\to0.5$ is
on the total information length ${\cal L}_\infty$. In figure \ref{Fig9} these
values are 25 and 16, respectively, whereas in figure \ref{Fig10} they are 35
and 9.5. That is, in both cases decreasing $\gamma$ yields larger
${\cal L}_\infty$ values than increasing $\gamma$ does, consistent with the
peaks being narrower, and hence passing through more statistically
distinguishable states. Next, comparing 25 for $\gamma=0.5\to0.1$ versus 35 for
$\gamma=0.5\to0$, this is exactly as one might expect: having the peak travel
somewhat further yields extra information length. However, comparing 16 for
$\gamma=0.1\to0.5$ versus 9.5 for $\gamma=0\to0.5$ is puzzling then! The peak
has further to travel, but accomplishes it with less information length. The
reason is precisely this extreme broadening, which substantially reduces the
number of distinguishable states along the way. See also \cite{KH17,Entropy},
where the same effect was studied for Gaussian PDFs, and values of $D$ as
small as $10^{-7}$, leading to fundamentally different scalings of
${\cal L}_\infty$ with $D$ for inward and outward processes.\\

Returning to the central question of this paper, namely how close the
time-dependent PDFs are to gamma distributions, the results for figure
\ref{Fig9} are similar to the previous ones. In particular, we recall that
before in figure \ref{Fig5} we had the difference scaling as $D^{1/2}$, so a
smaller $D$ here means a smaller difference. These results are approaching the
small $D$ regime where gamma distributions become very close to Gaussians
anyway, which generally remain close to Gaussian as they move.\\

However, for the $\gamma=0\to0.5$ process in figure \ref{Fig10}, the intermediate
stages do not look much like gamma distributions. (The final equilibrium is
indistinguishable from a gamma distribution though, consistent with $Q$ being
completely negligible for these values of $x$.) For the intermediate stages,
these were found to be so different from gamma distributions that attempting to
fit a gamma distribution having the same $\langle x\rangle$ and $\sigma$ made
little sense; this extreme broadening and long tail trailing behind the peak
meant that both $\langle x\rangle$ and $\sigma$ were too different from the
normal expectation that they should be measures of `peak' and `width'.\\

\begin{figure}
\centering
\includegraphics[scale=0.9]{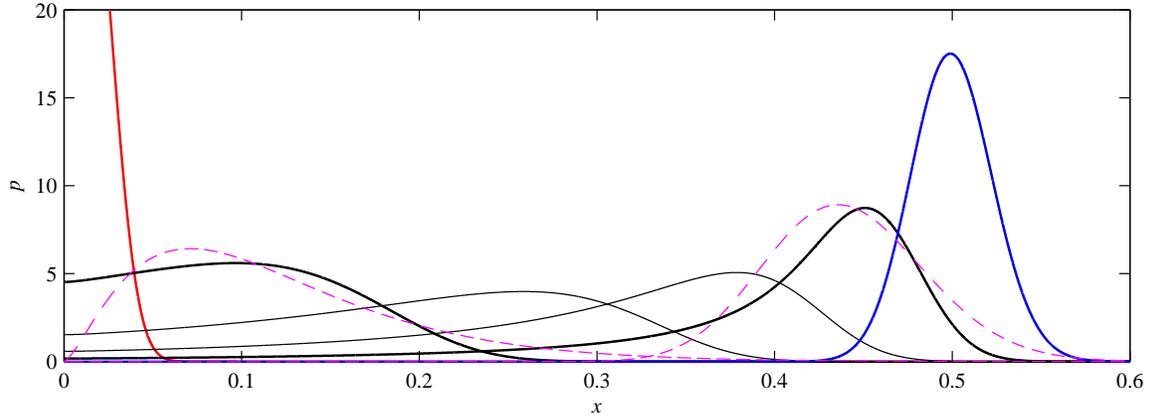}
\caption{The $\gamma=0\to0.5$ process as in figure \ref{Fig10}, but now shown
in more detail. The dashed (magenta) curves are the gamma distributions that
best fit the two thicker curves at intermediate times. Note how even a
`best-fit' is a rather poor approximation to the actual PDFs.}
\vskip 1cm
\label{Fig11}
\end{figure}

Instead, we simply asked the question, which values of $a$ and $b$ would minimize
the quantity $\int|p - p_{\rm bf}|\,dx$, where $p$ is the time-dependent PDF
to be fitted, and $p_{\rm bf}$ is the best-fit gamma distribution. Unlike our
previous difference formula, this does not yield simple analytic formulas for the
$a$ and $b$ to choose, but is numerically still straightforward to implement.
Figure \ref{Fig11} shows the results, for two of the intermediate stages in the
$\gamma=0\to0.5$ process. We can see that the fit is rather poor, indicating
that these PDFs are {\it significantly} different from gamma distributions.\\

This misfit is also not caused by the inclusion of $Q$; if this or any similar
central peak is evolved for either small or zero $Q$ in the Fokker-Planck
equation, the result is always similar to here. As explained also in
\cite{KH17,Entropy}, the dynamics of how central peaks move away from the origin
is simply different from how peaks already away from the origin move, regardless
of whether the final states are Gaussians as in \cite{KH17,Entropy}, or gamma
distributions as here.\\

\section{Conclusion}

Gamma distributions are among the most popular choices for modelling a broad
range of experimentally determined PDFs. It is often assumed that time-dependent
PDFs can then simply be modelled as gamma distributions with time-varying
parameters $a$ and $b$. In this work we have demonstrated that one should be
cautious with such an approach. By numerically solving the full time-dependent
Fokker-Planck equation, we found that there are three sets of circumstances
where the PDFs can differ significantly from gamma distributions:
\begin{itemize}
\item
If $D<\gamma$, so that stationary solutions exist, but $D$ is also sufficiently
close to $\gamma$ that a gamma distribution differs significantly from a
Gaussian, then the time-dependent PDFs will also differ significantly from
gamma distributions.
\item
If $D>\gamma$, stationary gamma distributions do not exist at all. Instead,
peaks move ever closer to the origin, and in the process increasingly differ
from gamma distributions.
\item
If the initial condition is a peak right on the origin -- either as a result of
adding additive noise to produce stationary solutions even for $D>\gamma$, or
simply as an arbitrary initial condition -- then any evolution away from the
origin will differ significantly from gamma distributions. Unlike the previous
two items, which become more pronounced for larger $D$, this effect is most
clearly visible for smaller $D$, where the mismatch between the naturally
narrower peaks and the extreme broadening seen in figure \ref{Fig11} becomes
increasingly significant.
\end{itemize}
{\bf
In summary, our results show that a simple Langevin equation model 
mimics the strong fluctuations of far-from-equilibrium systems.
This model has gamma distributions as steady-state solutions, 
but the time-dependent solutions can deviate considerably
from this law. This makes tasks such as Bayesian and frequentist 
inference of the model from data more complicated. On the other hand,
the model shows complex asymptotic dynamics with situations when 
a steady state is reached or not, different from the
one of the deterministic logistic model that invariably
evolves to the maximum capacity. The studied model 
is general enough and can be applied to many practical situations in biology, economics, finance, and physics.
}
Future work will apply some of these ideas to fitting actual data.\\
%

\appendix

\section{Derivation of the Fokker-Planck Equations}
In order to derive the Fokker-Planck equation (\ref{eq4}) from the Langevin
equation (\ref{eq2}), it is useful to introduce a generating function $Z$:
\begin{equation}
Z = e^{i \lambda x(t)}.
\label{b1}
\end{equation}
Then, by definition of `average', the average of $Z$ is related to the PDF,
$p(x,t)$, as
\begin{equation}
\langle Z \rangle = \int dx \,Z\, p(x,t)
 = \int dx \, e^{i \lambda x(t)}\, p(x,t).
\label{b2}
\end{equation}
Thus, we see that $ \langle Z \rangle$ is the Fourier transform of $p(x,t)$.
The inverse Fourier transform of $\langle Z \rangle$ then gives $p(x,t)$: 
\begin{equation}
p(x,t) = \frac{1}{2\pi} \int d\lambda  \, e^{-i \lambda x}\, \langle Z \rangle.
\label{b3}
\end{equation}
We note that Eq.\ (\ref{b3}) can be written as
\begin{equation}
p(x,t) = \left \langle \frac{1}{2 \pi} \int d\lambda
  \, e^{i \lambda (x- x(t))}\right \rangle
 = \left \langle \delta (x-x(t)) \right \rangle,
\label{b4}
\end{equation}
which is another form of $p(x,t)$. To obtain the equation for $p(x,t)$, we
first derive the equation for $\langle x \rangle$ and then take the inverse
Fourier transform as summarised in the following.\\

We differentiate $Z$ with respect to time $t$ and use Eq.\ (\ref{eq2}) to
obtain
\begin{equation}
\partial_t Z = i \lambda \partial_t x Z 
= i \lambda (\gamma x - \epsilon x^2 + \xi(t) x)Z
= \lambda \left [\gamma \partial_\lambda + i \epsilon \partial_{\lambda \lambda}
 + \xi \partial_\lambda \right] Z,
\label{b5}
\end{equation}
where $x Z = - i \partial_\lambda Z$ was used. The formal solution to
Eq.\ (\ref{b5}) is
\begin{equation}
Z(t) = \lambda \int dt_1\,  \left [\gamma \partial_\lambda + i \epsilon
 \partial_{\lambda \lambda} + \xi(t_1) \partial_\lambda \right] Z(t_1).
\label{b6}
\end{equation}
The average of Eq.\ (\ref{b5}) gives
\begin{equation}
\partial_t \langle Z \rangle  
= \lambda (\gamma \partial_\lambda + i \epsilon \partial_{\lambda \lambda} )
 \langle Z \rangle + \lambda \langle \xi(t) \partial_\lambda Z (t) \rangle.
\label{b7}
\end{equation}
To find $\langle \xi(t) \partial_\lambda Z (t) \rangle$, we use Eq.\ (\ref{b6}) iteratively as
follows:
\begin{eqnarray}
\langle \xi(t) \partial_\lambda Z  \rangle &=&
\left \langle \xi(t) \partial_\lambda \left [ \lambda \int dt_1\,
 \left [\gamma \partial_\lambda + i \epsilon \partial_{\lambda \lambda}
 + \xi(t_1) \partial_\lambda \right] Z(t_1)\right ] \right\rangle
\nonumber \\
&=& 
\langle \xi(t) \rangle \partial_\lambda \left [ 
\lambda \int dt_1\, \left [\gamma \partial_\lambda + i \epsilon
 \partial_{\lambda \lambda}\right] \langle Z(t_1)\rangle \right]
+ \partial_\lambda  \left [\lambda  \int dt_1 \, \langle \xi(t) \xi(t_1)
 \rangle \partial_\lambda \langle Z(t_1) \rangle \right ]  
\nonumber \\
&=& \partial_\lambda \Bigl[ \lambda \left[ D \partial_\lambda \langle Z(t)
 \rangle \right]\Bigr].
\label{b8}
\end{eqnarray}
Here we used the independence of $\xi(t)$ and $Z(t_1)$ for $t_1 <  t$,
$\langle \xi(t) Z(t_1) \rangle  = \langle \xi(t) \rangle \langle Z(t_1)
\rangle = 0$, together with Eq.\ (\ref{eq3}), $\int_{0}^{t} dt_{1}\,
\delta (t-t_{1}) = {1}/{2}$, and  $\langle \xi \rangle = 0$.
By substituting Eq.\ (\ref{b8}) into Eq.\ (\ref{b7}) we obtain
\begin{equation}
\partial_t \langle Z \rangle  =  \lambda (\gamma \partial_\lambda
 + i \epsilon \partial_{\lambda \lambda} ) \langle Z \rangle +
 \lambda \partial_\lambda \Bigl[ \lambda \left[ D \partial_\lambda
 \langle Z(t) \rangle \right]\Bigr].
\label{b9}
\end{equation}
The inverse Fourier transform of Eq.\ (\ref{b9}) then gives us 
\begin{equation}
\frac{\p}{\p t}\,p(x,t)
= - \frac{\p}{\p x}\Bigl[(\gamma x -\epsilon x^2)p(x,t)\Bigr]
+D\,\frac{\p}{\p x}\left[x\,\frac{\p}{\p x}\Bigl[x\,p(x,t)\Bigr]\right]
\label{b10}
\end{equation}
which is Eq.\ (\ref{eq4}). Specifically, the inverse Fourier transforms of the
first and last terms in Eq.\ (\ref{b9}) are shown explicitly in the following:
\begin{eqnarray}
&&\frac{1}{2 \pi} \int d\lambda \, e^{-i \lambda x} \langle \partial_t Z \rangle
= \partial_t \left[ \frac{1}{2 \pi} \int d\lambda \, e^{-i \lambda x}
 \langle Z \rangle\right]  =\frac{\p}{\p t}\,p(x,t),
\label{b11} \\
&& \frac{D}{2 \pi} \int d\lambda \, e^{-i \lambda x} \lambda \partial_\lambda
 \left [ \lambda  \partial _ \lambda \langle  Z \rangle \right]
=D\,\frac{\p}{\p x}\left[x\,\frac{\p}{\p x}\Bigl[x\,p(x,t)\Bigr]\right],
\label{b12}
\end{eqnarray}
where integration by parts was used twice in obtaining Eq.\ (\ref{b12}).
The additional $Q\partial_{xx}p$ term in the Fokker-Planck equation
(\ref{FP2}) can be derived in the same way.\\

\section{Time-dependent Analytical Solutions of Eq.\ (\ref{eq4})}
We begin by making the change of variables $y=1/x$ in Eq.\ (\ref{eq2})
to obtain
\begin{equation}
\frac{dy}{dt} = -(\gamma + \xi) y + \epsilon.
\label{eq100}
\end{equation}
By using the Stratonovich calculus \cite{Klebaner,Gardiner,WongZakai}, the
solution to Eq.\ (\ref{eq100}) is found as
\begin{equation}
y(t) =  y_0 e^{-(\gamma t + B(t))}
+ \epsilon e^{-(\gamma t + B(t))} \int_0^t dt_1 e^{(\gamma t_1 + B(t_1))},
\label{eq101}
\end{equation}
where $y_0 = y(t=0)$ and $B(t) = \int_0^t\, dt_1\, \xi(t_1)$ is the
Brownian motion. Therefore,
\begin{equation}
x(t) =  \frac{x_0 e^{\gamma t + B(t)}}
{1 + \epsilon x_0  \int_0^t dt_1 e^{(\gamma t_1 + B(t_1))}},
\label{eq102}
\end{equation}
where $x_0 = x(t=0)$. In Eq.\ (\ref{eq102}), $e^{B(t)}$ is the geometric
Brownian motion while  $e^{-\gamma t - B(t)}$ is the geometric Brownian
motion with a drift (e.g.\ \cite{Klebaner}). The time integral of the latter
is used in understanding stochastic processes in financial mathematics and
many other areas \cite{bertoin,matsumoto}. In particular, in the long time
limit, its PDF can be shown to be a gamma distribution. However, this PDF of
$x$ is not particularly useful as it involves complicated summations and
integrals that cannot be evaluated in closed form \cite{bertoin,matsumoto}.\\

\end{document}